\documentclass[12pt,fleqn]{article}
\makeatletter
\@addtoreset{equation}{section}
\makeatother

\newcounter{saveeqn}
\newcommand {\bcs} {boundary conditions}

\newcommand {\rhs} {right-hand side}
\newcommand {\ASth} {Atiyah-Singer index theorem}
\newcommand {\ncl} {non-contractible loop}

\newcommand {\ewsm} {electroweak standard model}

\newcommand {\YMths} {Yang-Mills theories}

\newcommand {\YMHth} {Yang-Mills-Higgs theory}
\newcommand {\sph} {sphaleron}

\newcommand {\fes} {forward elastic scattering amplitude}

\newcommand {\Istar} {\rm I$^{\star }$}
\newcommand {\Ibar} {$\bar{{\rm I}}$}
\newcommand {\IIbar} {\rm I$\, \bar{{\rm I}}$}

\newcommand {\Sstar} {\rm S$^{\star }$}
\newcommand {\dstar} {$d^{\star }$}
\newcommand {\Dstar} {$D^{\star }$}

\newcommand {\mIstar} {{\rm I}^{\star }}  
\newcommand {\mI}    {{\rm I}}
\newcommand {\mIbar} {\bar{{\rm I}}}
\newcommand {\mIIbar} {{\rm I}\,\bar{\rm I}}
\newcommand {\mdstar} {d^{\star}}
\newcommand {\mDstar} {D^{\star}}
\newcommand {\mtildeESstar} {\tilde{E}_{ {\rm S}^{\star} } }

\newcommand {\rmd}    {{\rm d}}

\newcommand {\gsim}{\mathrel{\hbox{\rlap{\lower.55ex \hbox {$\sim$}}            
            \kern-.3em \raise.4ex \hbox{$>$}}}}
\newcommand {\lsim}{\mathrel{\hbox{\rlap{\lower.55ex \hbox {$\sim$}}            
            \kern-.3em \raise.4ex \hbox{$<$}}}}
\newcommand {\vectwo}[2] {\left(\begin{array}{c}#1\\#2\end{array}\right)}

\newcommand {\beq} {\begin{equation}}
\newcommand {\eeq} {\end{equation}}

\def\u  {{\bf u}}
\def\v  {{\bf v}}
\def\w  {{\bf w}}

\def\id{{\rm 1\kern-.12em
\rule{0.3pt}{1.5ex}\raisebox{0.0ex}{\rule{0.1em}{0.3pt}}}}

\def\C{{\rm\kern.24em
   \vrule width.02em
       height1.4ex depth-.05ex
   \kern-.26em C}}

\def\R  {{\rm I\kern-.15em R}}

\def\L  {{\rm I\kern-.25em L}}

\def\N{{\rm I\kern-.23em N}}

\begin{document}

\begin{titlepage}
\noindent hep-th/9709194 \hspace*{\fill} KA--TP--15--1997 \\[0.5ex] 
Nucl. Phys. B 517 (1998) 142
\begin{center}
\vspace{3\baselineskip}
{\Large \bf Fermion zero-modes of a new constrained instanton\\[1ex]
             in \YMHth}\\
\vspace{2\baselineskip}
{\large F. R. Klinkhamer} \footnote{E-mail:
                           frans.klinkhamer@physik.uni-karlsruhe.de}\\
\vspace{1\baselineskip}
 Institut f\"ur Theoretische Physik\\ Universit\"at Karlsruhe\\
 D--76128 Karlsruhe\\Germany\\
\vspace{3\baselineskip}
{\bf Abstract} \\
\end{center}
{\small
\noindent Self-consistent \emph{Ans\"{a}tze}
are presented for the left- and right-handed isodoublet fermion zero-modes
of the constrained instanton \Istar~in the vacuum sector of euclidean
$SU(2)$ Yang-Mills-Higgs theory.
These left- and right-handed fermion wave functions do not coincide and,
most likely, have maxima at different positions.
This may be important for the fermion zero-mode
contribution to the euclidean 4-point Green's function in chiral
Yang-Mills-Higgs theory and
the high-energy behaviour of fermion-fermion scattering processes.
}
\vspace{2\baselineskip}
\begin{tabbing}
PACS \hspace{1.5em} \= : \hspace{0.5em} \=
                            11.27.+d; 03.65.Pm; 11.15.-q \\[0.5ex]
Keywords            \> : \> Instanton; Fermion zero-mode; Scattering \\
\end{tabbing}
\end{titlepage}

\section{Introduction}

Recently, a self-consistent \emph{Ansatz} has been presented \cite{KW96}
for a new constrained instanton (finite action, localized, 4-dimensional
euclidean solution) in the topological charge-zero sector of $SU(2)$ \YMHth.
A numerical approximation to the bosonic fields of this instanton
\Istar~has been obtained previously \cite{K93a},
but the complete numerical solution
for the profile functions of the \emph{Ansatz} is not available yet.
In this paper, we consider the response of massless, chiral fermions
to the generic bosonic fields of \Istar, with the fermions in the fundamental
(isodoublet) representation of the $SU(2)$ gauge group.
The appropriate \emph{Ans\"{a}tze}
for the fermion zero-modes  are constructed and some partial
solutions of
the reduced Weyl equations are obtained analytically.

    The outline of this paper is then as follows. In Section 2, the euclidean
chiral $SU(2)$ \YMHth~considered is reviewed. In Section 3,
the bosonic fields of the instanton \Istar~are discussed, together with
the relevant symmetries. In Section 4, these symmetries are used to
construct
the \emph{Ans\"{a}tze} for the left- and right-handed isodoublet fermion
zero-modes. In Section 5, the two Weyl equations are evaluated, which
prove the \emph{Ans\"{a}tze} to be self-consistent. Also, in Section 5, the
related Dirac eigenvalue equation is established.
The Dirac eigenvalues display the phenomenon of
spectral flow, when evaluated over a \ncl~of bosonic field
configurations passing through the classical vacuum and \Istar~\cite{W82,K92}.
Two of these Dirac eigenvalues are expected to cross at \Istar~and give a pair of
normalizable fermion zero-modes, one of each chirality.
In Section 6, some partial solutions of the reduced Weyl equations
in the \Istar~background  are established analytically.
An approximate solution by courtesy of supersymmetry is also presented.  
The most important result, though, is that the left- and right-handed fermion
zero-modes are shifted with respect to one another. This traces back to
the conjectured
structure of the bosonic fields of \Istar, where two cores (with Higgs
field vanishing) are kept apart in a balancing act between the repulsive
Yang-Mills force and the attractive Higgs force, the  core separation
\dstar~being of order $M_W^{-1}$ (the Yukawa range of the
massive $W$ vector bosons). Here, and in the following,
natural units are used with $\hbar$ $=$ $c$ $=$ $1$.

The main focus of the present paper is on the fermion solutions
themselves, but in Section 7 we mention one possible application of
the \Istar~fermion zero-modes in the chiral $SU(2)$  \YMHth~with
\emph{two} flavors of massless left-handed fermions, namely their
contribution to the euclidean 4-point Green's function
relevant to the forward elastic scattering amplitude.
The relative shift of the two types of fermion
zero-modes gives a non-trivial phase factor in the Fourier transformed
euclidean 4-point function. In Section 8, finally, we argue that this
phase factor, analytically continued to Minkowski space-time,
may signal unusual behaviour
at center-of-mass scattering energies of
order $M_W / \alpha_w$, where $\alpha_w$ $\equiv$ $g^2 / 4\pi$
stands for the fine-structure constant of the $SU(2)$ gauge fields.

\section{Theory}

    In this section we briefly review the theory considered in this paper,
mainly in order to establish notation. The theory considered can be viewed
as a simplified version of the \ewsm, with $SU(2)$ gauge fields, a single
isodoublet of complex scalars and
$N_f$ isodoublets of massless left-handed Weyl fermions. The euclidean
action for this chiral $SU(2)$ \YMHth~\cite{YM54,H66} is given by
\beq
 A_{\mbox {\rm \tiny YMH}} = \int\limits_{\R^4} \rmd^4x   \left[ \;
 \frac{1}{4} \left(W^{a}_{\mu \nu}\right)^2
 + |D_\mu \Phi|^2  + \lambda \left( |\Phi|^2 - \frac{v^2}{2} \right)^2
 + \sum_{f=1}^{N_{f}} \Psi_f^{\dagger}\,i\, \bar{\sigma}^{\mu} D_{\mu}\Psi_f
 \; \right]\; ,
\label{eq:AYMH}
\eeq
with
\begin{eqnarray*}
D_\mu \; \cdot &\equiv&
      \left(\,\partial_\mu + g\,W_\mu \, \right)\;\cdot\;\;,\\[.5ex]
W_{\mu\nu}&\equiv& W_{\mu\nu}^a \,T^a \equiv
                   \partial_\mu W_\nu -\partial_\nu W_\mu+
                   g\,\left[\,W_\mu\, ,W_\nu\,\right]\;,\\
W_{\mu \phantom{\nu}}
          &\equiv& W_{\mu \phantom{\nu}}^a \,T^a \equiv
                   W_\mu^a \,\frac{\tau^a}{2\,i} \;\;.
\end{eqnarray*}
Here, the $\tau^a$, $a = 1$, $2$, $3$,
denote the ``weak isospin'' Pauli matrices
\[
  \tau^1 \equiv \left( \begin{array}{cc} 0 & 1 \\ 1 & 0 \end{array}\right)
  \;, \qquad
  \tau^2 \equiv \left( \begin{array}{cc} 0 & -i \\ i & 0 \end{array}\right)
  \;, \qquad
  \tau^3 \equiv \left( \begin{array}{cc} 1 & 0 \\ 0 & -1 \end{array}\right)\;,
\]
and the $\sigma^\mu$,  $\bar{\sigma}^\mu$, $\mu = 1$, $2$, $3$, $4$,
the spin matrices
\[
    \sigma^\mu \equiv (\, i\sigma^1, i\sigma^2, i\sigma^3, \id\,)\; , \quad
    \bar{\sigma}^\mu \equiv ( -i\sigma^1, -i\sigma^2, -i\sigma^3, \id\,) =
    \sigma^{\mu\, \dagger} \;,
\]
where the matrices $\sigma^a$ have the same entries as the $\tau^a$ above.
The chiral fermions are massless, whereas the  three vector bosons $W^a$
have mass $M_W \equiv \frac{1}{2}\,g\,v$  and the single Higgs scalar
$H$ mass $M_H \equiv \sqrt{2\,\lambda}\, v$.
Except for the last two sections of this paper, we
always consider a single flavor of chiral fermions ($N_f=1$) and drop the
flavor index $f$. 

In order to obtain non-singular instanton solutions for the bosonic fields,
the action $A_{\mbox {\rm \tiny YMH}}$
has to be supplemented by
a constraint term $A_{\mbox {\rm \tiny C}}$ \cite{A81},
which is the 4-dimensional integral of an appropriate
local operator $O_d$ of canonical mass dimension $d>4$ (for example,
${\rm Tr}\, (W_{\mu\nu})^4$  with $d=8$).
See Ref. \cite{K93a} for how this works in practice.

    For later use, we introduce some further spinor conventions, which
follow basically those of Ref. \cite{JR77}. The euclidean Dirac matrices
$\gamma^\mu$, $\mu = 1$, $2$, $3$, $4$, have anticommutation relations
\[
   \{ \gamma^\mu,  \gamma^\nu \} = 2 \, \delta^{\mu\nu}
\]
and can be represented by the following hermitian $4 \times 4$ matrices:
\beq
  \gamma^\mu =\left(\begin{array}{cc} 0           & \bar{\sigma}^\mu  \\
                                      \sigma^\mu  & 0 \end{array} \right)\; ,
\label{eq:chiralrepr}
\eeq
in terms of the $2 \times 2$ matrices $\sigma^\mu$,  $\bar{\sigma}^\mu$
given above.  Defining the chirality operator
\[
  \gamma^5 \equiv \gamma^1 \gamma^2 \gamma^3 \gamma^4  =
  \left(\begin{array}{cc} \id  &  0 \\
                          0    & -\id     \end{array} \right)
\]
and the corresponding projection operators
\[
   P_L \equiv \frac{1}{2}\, \left( \, \id - \gamma^5 \,\right) \;, \quad
   P_R \equiv \frac{1}{2}\, \left( \, \id + \gamma^5 \,\right) \;,
\]
a general Dirac spinor $\Psi_{\rm D}$ can be written as the composite of
two Weyl spinors
\beq
   \Psi_{\rm D} = \left(\begin{array}{c}
                  \Psi_{R\,\alpha}  \\  \Psi_{L}^{\phantom{L}\,\dot{\alpha}}
                  \end{array} \right)\; ,
\eeq
with indices $\alpha$ $=$ $1$, $2$, and $\dot{\alpha}$ $=$ $1$, $2$
(van der Waerden notation).
For the spin matrices this gives the following index structure:
\[
    \sigma^{\mu\,\dot{\alpha}\alpha}\; , \quad
    \bar{\sigma}^\mu_{\phantom{\mu}\,\alpha\dot{\alpha}} \;\;,
\]
and we refer to the Appendix A of Ref. \cite{WB83} for further details.

    The Dirac eigenvalue equation
\beq
     i \, \gamma^\mu D_\mu \, \Psi_{\rm D} = \epsilon \,\Psi_{\rm D}
\label{eq:Dirac}
\eeq
now reduces to two coupled equations for the Weyl spinors $\Psi_L$ and
$\Psi_R$. For vanishing eigenvalue $\epsilon $,
these equations decouple, giving the two fundamental Weyl equations
\beq
     i \, \bar{\sigma}^\mu D_\mu^{\;(L)}\, \Psi_L = 0 \; , \quad
     i \,      \sigma^\mu  D_\mu^{\;(R)}\, \Psi_R = 0 \; .
\label{eq:Weyl}
\eeq
The suffixes ($L$) and ($R$) on the covariant derivatives in
(\ref{eq:Weyl}) are to remind
us that for \emph{massless} fermions the gauge group representations
of $\Psi_L$ and $\Psi_R$ may differ \cite{PS95},
which will be important in Sect. 5.

\section{Bosonic fields}

    The fields considered in this paper have an axial $SO(2)$ symmetry,
which makes it useful to introduce cylindrical coordinates defined
in terms of the cartesian coordinates by
\[
(\rho\cos{\varphi},\rho\sin{\varphi},z,\tau) \equiv (x^1,x^2,x^3,x^4)\;,
\]
and a triad of isospin matrices
\[
     \u \equiv   \cos{\varphi}\;T^1 + \sin{\varphi}\; T^2 \;,\quad
     \v \equiv - \sin{\varphi}\;T^1 + \cos{\varphi}\; T^2 \;,\quad
     \w \equiv   \,T^3                                    \;,
\]
defined in terms of the isospin $\frac{1}{2}$ representation
$T^a \equiv \tau^a/(2\,i)$.

   The bosonic fields of the constrained instanton \Istar~are then given by
\begin{eqnarray}
 g\,W_\rho   &=& \frac{1}{\rho} \: \left\{ \:
              C_{1\phantom{0}} \, \u +  C_{2\phantom{0}}  \, \v +
              C_{6\phantom{0}}  \, \w \:\right\} \;,
\quad
 g\,W_\varphi =  \frac{1}{\rho} \: \left\{ \:
              C_{3\phantom{0}} \,  \u + C_{4\phantom{0}} \, \v +
              C_{5\phantom{0}} \, \w \: \right\} \;,
\nonumber \\
 g\,W_z      &=& \frac{1}{z} \: \left\{ \:
              C_{7\phantom{0}} \, \u + C_{8\phantom{0}} \, \v +
              C_{9\phantom{0}} \, \w \:  \right\} \;,
\quad
 g\,W_\tau    =  \frac{1}{\tau} \: \left\{ \:
              C_{10} \, \u + C_{11} \, \v + C_{12}\,\w \:\right\} \;,
\nonumber \\[1ex]
 \Phi     &=& \frac{v}{\sqrt{2}} \left(
 \begin{array}{c}(\,iH_1+ H_2\,)\,e^{-i\varphi}\\iH_3 + H_4\end{array}
                                 \right)\;,
\label{eq:axialfields}
\end{eqnarray}
where the signs of the coefficient functions
$C_5$, $C_8$, $C_{11}$, $H_1$ and $H_2$ have been changed
compared to Ref. \cite{KW96} and $H_4$ has replaced $H_0$.   An
important characteristic of \Istar~is a certain discrete
symmetry ${\rm D}_1$ (to be discussed further in the next section),
which implies for the real coefficient functions
$C_i$ and $H_j$ the following reflection properties:
\begin{eqnarray}
C_i(\rho,z,\tau)&=&+\;C_i(\rho,-z,\tau)\;,\quad\; i=1,4,5,7,10\;,
\nonumber\\
C_i(\rho,z,\tau)&=&-\;C_i(\rho,-z,\tau)\;,\quad\; i=2,3,6,8,9,11,12\;,
\nonumber\\
H_j(\rho,z,\tau)&=&-\;H_j(\rho,-z,\tau)\;,\quad   j=1,4 \;,
\nonumber\\
H_j(\rho,z,\tau)&=&+\;H_j(\rho,-z,\tau)\;,\quad   j=2,3 \; .
\label{eq:CHreflsymm}
\end{eqnarray}
Further \bcs~on the coefficient functions, in particular those at infinity,
can be found in Ref. \cite{KW96}.
As mentioned before, we have at this moment only an
approximation \cite{K93a} to the behavior of the
coefficient functions $C_i(\rho,z,\tau)$,
$H_j(\rho,z,\tau)$ and not the full numerical solution
of the reduced field equations.

    This completes our brief discussion of the constrained instanton
\Istar. For later use, we also give the bosonic fields of the
well-known instanton 
${\rm I}_{\rm BPSTH}$ \cite{BPST75,tH76b}, or I for short,
\beq
g\,W_\mu  = - f(x) \;  \partial_\mu \,U \; U^{-1}   \nonumber \;,\quad
\Phi   = \frac{v}{\sqrt{2}}\; h(x) \; U \vectwo{0}{1}\nonumber \;,\quad
U      = \hat{x}^\mu \,  \tau^\mu \; ,
\label{eq:Ifields}
\eeq
with the radial coordinate $x$ $\equiv$ $\sqrt{x^\mu\, x^\mu}$,
the unit-vector  $\hat{x}^{\mu}$ $\equiv$ $x^{\mu} / x$ and the isospin
matrices  $\tau^{\mu}$ $\equiv$ $(i \vec{\tau}, \id\,)$.
These fields have $SO(4)$ rotation invariance and can certainly be
brought to the axial form (\ref{eq:axialfields}).
Moreover, the radial profile functions $f$ and $h$ are known analytically
in the limit $\tilde{\rho}\,v \to 0$
\beq
f(x)  =  h(x)^2 = \frac{x^{2}}{x^{2}+\tilde{\rho}^{\,2}} \;\;,
\label{eq:Ifunctions}
\eeq
with $\tilde{\rho}$ the scale parameter of the solution as fixed by the
constraint
(a tilde distinguishes this scale from the axial coordinate $\rho$).
Note that the Higgs field vanishes at the origin, independently
of the gauge, and that the BPST gauge fields (\ref{eq:Ifields}),
(\ref{eq:Ifunctions}) are self-dual
\beq
 \phantom{.}^\star W_{\kappa\lambda} \equiv
 \frac{1}{2}\,\epsilon_{\kappa\lambda\mu\nu}\;W_{\mu\nu} =
 +\, W_{\kappa\lambda} \; ,
\label{eq:selfdual}
\eeq
with $\epsilon_{\kappa\lambda\mu\nu}$ the completely antisymmetric
pseudotensor ($\epsilon_{1234}=1$). For the anti-instanton \Ibar~the
matrix $U$ in (\ref{eq:Ifields}) is replaced by $U^\dagger$ and
the resulting gauge fields are
anti-self-dual, with a minus sign on the \rhs~of (\ref{eq:selfdual}).
The (anti-)self-duality property of the contrained BPSTH (anti-)instanton
solution is lost for finite values of $\tilde{\rho}$ \cite{A81}.

    The topological charge (Pontryagin number)
\beq
Q \: \equiv \int\limits_{\R^4}\rmd^4x \: q \;
  \equiv \; \frac{1}{32\,\pi^2} \: \int\limits_{\R^4}\rmd^4x \:
   \phantom{.}^\star W_{\kappa\lambda}^a \, W_{\kappa\lambda}^a
\label{eq:Q}
\eeq
takes the value ($-1$) +1 for the BPSTH (anti-)instanton ($\,\mIbar\,$) I,
whereas the topological charge is 0 for \Istar. In fact, the instanton
\Istar~may be viewed as an unstable di-atomic molecule built
out of an approximate instanton I and anti-instanton \Ibar.
Note, however, that this picture has not yet been established
rigorously. It follows, instead, from the observation that the $z=0$
slice of \Istar~corresponds to the static (time-independent)
sphaleron \Sstar~\cite{K93b}, for
which a numerical solution of the field equations has indeed
established the molecule-like structure (and not the ring-like
structure allowed in principle; see Fig. 1 of Ref. \cite{KW96} 
for a sketch of the two alternatives). Henceforth, we take the
molecule-like structure of \Istar~for granted, with a distance
$\mdstar$ $=$ ${\rm O}(M_W^{-1})$ between the two Higgs zeros
on the $\tau$-axis.

\section{Fermion Ans\"{a}tze}

    The \emph{Ans\"{a}tze} for the fermionic fields follow directly from
the continuous and discrete symmetries of \Istar.
The axial $SO(2)$ symmetry for an isodoublet of left-handed Weyl fermions
is generated by the operator
\beq
     K_3 \equiv -i\, \partial_\varphi +
         \frac{\sigma^3}{2} + \frac{\tau^3}{2} \left( P_L - P_R \right) \; ,
\label{eq:K3}
\eeq
with the projection operators $P_L$ and $P_R$ taking the values
1 and 0, respectively. Introducing the kets $|L, s_3, t_3 \rangle$,
with $s_3$ and $t_3$ the eigenvalues of $\sigma^3/2$ and $\tau^3/2$,
respectively, the four $K_3=0$ eigenvectors are
\footnote{~These eigenvectors also appear in a different context, namely
the static $Z$-string solution of the \ewsm.
There, the four eigenvectors combine
into pairs by the action of an additional reflection symmetry ${\rm R}_1$,
which is due to the 2-dimensional nature of the fields \cite{KR97}.}
\beq
      | L, 1/2, -1/2\rangle\; , \;\;  |L, -1/2, 1/2\rangle\; , \;\;
      e^{+i \varphi}\,|L, -1/2, -1/2\rangle               \; , \;\;
      e^{-i \varphi}\,|L, 1/2, 1/2\rangle                 \; .
\label{eq:Leigenkets}
\eeq
Using the column vector notation
\[
    \left( \begin{array}{c}
                           \left( \begin{array}{c} \times \\ \times
                           \end{array} \right)_{\rm isospin}
    \\
                           \left( \begin{array}{c} \times \\ \times
                           \end{array} \right)_{\rm isospin}
    \end{array} \right)_{\rm spin} \;,
\]
the $K_3=0$ eigenvectors (\ref{eq:Leigenkets}) can be written as
\beq
\left(\begin{array}{c} 0   \\ 1 \\ 0  \\ 0         \end{array}\right)\,,\quad
\left(\begin{array}{c} 0   \\ 0 \\ 1  \\ 0         \end{array}\right)\,,\quad
\left(\begin{array}{c} 0 \\ 0 \\ 0 \\ e^{+i\varphi}\end{array}\right)\,,\quad
\left(\begin{array}{c} e^{-i\varphi}\\ 0 \\ 0 \\ 0 \end{array}\right)\,.
\eeq
The general \emph{Ansatz} for an axisymmetric left-handed Weyl isodoublet
is then
\beq
     \Psi_L =      \left(\begin{array}{c}
                         F_{L\,\uparrow}   \; e^{-i\varphi}\\
                         G_{L\,\uparrow}                   \\
                         G_{L\,\downarrow}                 \\
                         F_{L\,\downarrow} \; e^{+i\varphi}
                   \end{array} \right) \; ,
\label{eq:PsiL}
\eeq
in terms of four complex functions of the axial coordinates $\rho$,
$z$ and $\tau$ (of course, $F_{L\,\uparrow}$ and $F_{L\,\downarrow}$
must be proportional to $\rho$ for continuity of the fields).

    We now turn to the role of the discrete symmetry of \Istar.
As discussed in Ref. \cite{KW96}, this symmetry transformation
${\rm D}_1$ consists of three parts:
\begin{enumerate}
\item the coordinate transformation $\rho \to \rho$,
      $\varphi \to \pi - \varphi$,
      $z \to -z$, $\tau \to \tau$;
\item the complex conjugation of the fields $W$, $\Phi$, $\Psi$;
\item the global gauge transformation
      $W$ $\to$ $\Gamma \,W \,\Gamma^{-1}$,
      $\Phi $ $\to$ $\Gamma \,  \Phi$, $\Psi$ $\to$ $\Gamma \, \Psi$,
      with parameter $\Gamma=-\id\,$ in the center of the gauge group $SU(2)$.
\end{enumerate}
It is straightforward to check that the discrete symmetry
${\rm D}_1$ of the scalar isodoublet
\[
   \Phi = \left(\begin{array}{c}
                         F  \, e^{-i\varphi}\\
                         G
          \end{array} \right)
\]
gives back the structure (\ref{eq:axialfields}),
(\ref{eq:CHreflsymm}) of the \Istar~scalar fields. For the case of a fermion
isodoublet, the \emph{Ansatz} (\ref{eq:PsiL}) is restricted as follows
\begin{eqnarray}
F_{L\,\uparrow}&=& i \,\frac{\rho \,z}{X^2} \,k_1 + \frac{\rho}{X} \, k_2 \;,
\quad
G_{L\,\uparrow} =  i \,k_3 + \frac{z}{X}  \,k_4 \;,
\nonumber\\[0.5ex]
F_{L\,\downarrow}&=&
                 i \,\frac{\rho \,z}{X^2} \,l_1\: + \frac{\rho}{X}\, l_2\; ,\:
\quad
  G_{L\,\downarrow} =  i \,l_3\: + \frac{z}{X}  \,l_4 \;,
\label{eq:FGL}
\end{eqnarray}
in terms of eight real functions $k_j$ and $l_j\,$, $j=1$, $\cdots$ , $4$,
which are even in $z$
\beq
     k_j (\rho,z,\tau) = k_j (\rho,-z,\tau)\;, \quad
     l_j (\rho,z,\tau) = l_j (\rho,-z,\tau)\;,
\label{eq:klreflsymm}
\eeq
and the definition
\[
X^2 \equiv x^2+x_a^2 \equiv \rho^2 + z^2 + \tau^2 +x_a^2 \;,
\] 
for an arbitrary, but fixed, scale parameter $x_a$.
The regular functions $k_j$, $l_j$ have furthermore \bcs
\beq
    \lim_{|x| \to \infty} k_j =  \lim_{|x| \to \infty} l_j = 0
\label{eq:klbcs}
\eeq
and must give a normalizable spinor $\Psi_L$. This completes the \emph{Ansatz}
for the left-handed isodoublet fermion zero-mode in the \Istar~background.

The right-handed isodoublet fermion zero-mode is constructed in the same way.
The $K_3=0$ eigenvectors are now
\beq
      | R, 1/2, 1/2\rangle\; , \;\;  |R, -1/2, -1/2\rangle\; , \;\;
      e^{+i \varphi}\,|R, -1/2, 1/2\rangle \; , \;\;
      e^{-i \varphi}\,|R, 1/2, -1/2\rangle \; .
\label{eq:Reigenkets}
\eeq
Hence, the \emph{Ansatz} is
\beq
 \Psi_R =          \left(\begin{array}{c}
                         G_{R\,\uparrow}                  \\
                         F_{R\,\uparrow}   \; e^{-i\varphi}\\
                         F_{R\,\downarrow} \; e^{+i\varphi}\\
                         G_{R\,\downarrow}
                   \end{array} \right) \; ,
\label{eq:PsiR}
\eeq
with the following restrictions from the discrete symmetry ${\rm D}_1$:
\begin{eqnarray}
 F_{R\,\uparrow}&=& i \,\frac{\rho \,z}{X^2}\, m_1 +\frac{\rho}{X}\, m_2 \;,
\quad
 G_{R\,\uparrow} =  i \, m_3 + \frac{z}{X}  \, m_4 \;,
\nonumber\\[0.5ex]
 F_{R\,\downarrow}&=&i\,\frac{\rho \,z}{X^2}\, n_1\:+\frac{\rho}{X}\, n_2\;,\:
\quad
  G_{R\,\downarrow} =  i \, n_3\: + \frac{z}{X}  \, n_4 \;,
\label{eq:FGR}
\end{eqnarray}
where the real functions $m_j$ and $n_j$ obey the same conditions
(\ref{eq:klreflsymm}), (\ref{eq:klbcs})
as the functions $k_j$ and $l_j$.
Precisely which right-handed Weyl equation is solved by this \emph{Ansatz}
will be
explained in the next section.

For future reference, the appropriate \emph{Ans\"{a}tze} for
the fermion zero-modes of the 3-dimensional
sphaleron \Sstar~\cite{K93b} follow from the \Istar~\emph{Ans\"{a}tze}
by setting $z=0$ and implementing a further discrete symmetry ${\rm D}_2$
($z=0$ : $\rho \to \rho$, $\varphi$ $\to$ $\varphi + \pi$,
$\tau$ $\to$ $-\tau$).

\section{Weyl and Dirac equations}

Consider the left-handed Weyl equation corresponding
to the action (\ref{eq:AYMH})
\beq
     i \, \bar{\sigma}^\mu  D_\mu\, \Psi_L = \, i \;
     \left( \begin{array}{cc}
            D_\tau - i D_z & -e^{-i\varphi}(D_\varphi + i D_\rho) \\
            e^{i\varphi}(D_\varphi - i D_\rho) & D_\tau + i D_z \end{array}
     \right)
     \left( \begin{array}{c}
            \Psi_{L\,\uparrow} \\ \Psi_{L\,\downarrow}          \end{array}
     \right) = 0
\label{eq:WeylL}
\eeq
in the axisymmetric gauge field background (\ref{eq:axialfields}).
This can be evaluated readily for the fermion \emph{Ansatz} (\ref{eq:PsiL}),
with  the resulting four equations
\begin{eqnarray}
\lefteqn{
   \left[\, \left(\partial_\tau - \frac{C_9}{2\,z} \right) \mp
    i\left(\partial_z + \frac{C_{12}}{2\,\tau} \right) \,\right]
    F_{L} \!\!\mbox{\tiny$ \begin{array}{l} \uparrow
   \\[-.5ex] \downarrow \end{array} $}-
   \left[\, \frac{1}{2\,\rho} \left( C_1 - C_4 \right) \mp
    \frac{i}{2\,\rho}\left( C_2 + C_3 \right) \,\right]
    F_{L} \!\!\mbox{\tiny$ \begin{array}{l} \downarrow
   \\[-.5ex] \uparrow \end{array} $}-
}
\nonumber\\[1ex]
\lefteqn{
    \left[\, i \left( \partial_\rho - \frac{C_5}{2\,\rho} \right) \pm
    \frac{C_6}{2\,\rho} \,\right]
    G_{L} \!\!\mbox{\tiny$ \begin{array}{c} \downarrow
    \\[-.5ex] \uparrow \end{array} $}-
    \left[\, i \left( \frac{C_{10}}{2\,\tau} - \frac{C_8}{2\,z} \right) \pm
    \left( \frac{C_{11}}{2\,\tau} + \frac{C_7}{2\,z} \right) \,\right]
    G_{L} \!\!\mbox{\tiny$ \begin{array}{c} \uparrow
    \\[-.5ex] \downarrow \end{array} $}\! = 0 \;,
}
\nonumber\\[2ex]
\lefteqn{
    \left[\, \left(\partial_\tau + \frac{C_9}{2\,z} \right) \mp
    i\left(\partial_z - \frac{C_{12}}{2\,\tau} \right) \,\right]
    G_{L} \!\!\mbox{\tiny$ \begin{array}{c} \uparrow
\\[-.5ex] \downarrow \end{array} $}-
    \left[\, \frac{1}{2\,\rho} \left( C_1 + C_4 \right) \pm
    \frac{i}{2\,\rho}\left( C_2 - C_3 \right) \,\right]
    G_{L} \!\!\mbox{\tiny$ \begin{array}{c} \downarrow
\\[-.5ex] \uparrow \end{array} $}-
}
\nonumber\\[1ex]
\lefteqn{
  \left[\,i\left(\partial_\rho +\frac{1}{\rho}+\frac{C_5}{2\,\rho}\right)\mp
    \frac{C_6}{2\,\rho} \,\right]
    F_{L} \!\!\mbox{\tiny$ \begin{array}{c} \downarrow
    \\[-.5ex] \uparrow \end{array} $}-
    \left[\, i \left( \frac{C_{10}}{2\,\tau} + \frac{C_8}{2\,z} \right) \mp
    \left( \frac{C_{11}}{2\,\tau} - \frac{C_7}{2\,z} \right) \,\right]
    F_{L} \!\!\mbox{\tiny$ \begin{array}{c} \uparrow
    \\[-.5ex] \downarrow \end{array} $}\! = 0 \;,
}
\label{eq:FGLeqs}
\end{eqnarray}
where for the case of the \Istar~instanton the functions $F_L$ and $G_L$
have to be replaced by (\ref{eq:FGL}) and the gauge field functions $C_i$
by the appropriate expressions in terms of the functions $f_i$ of
Ref. \cite{KW96}.

    The result (\ref{eq:FGLeqs}) illustrates the axial symmetry
of the fermionic and bosonic fields, with the phases $\exp(\pm \,i\,\varphi)$
factoring out. Also, the $z$-reflection symmetry (\ref{eq:CHreflsymm}),
(\ref{eq:klreflsymm}) of \Istar~is respected by (\ref{eq:FGLeqs}), with
the complex structure
\beq
   {\rm Even} + i \, {\rm Odd} = 0
\eeq
for the first two equations and with the role of Even and Odd switched for the
last two. All in all, the \emph{Ansatz}
(\ref{eq:PsiL}), (\ref{eq:FGL}) has reduced the left-handed
Weyl equation (\ref{eq:WeylL}) in the \Istar~background
to eight real partial differential equations
(\ref{eq:FGLeqs})
for the eight real functions $k_j(\rho,z,\tau)$ and $l_j(\rho,z,\tau)$.
In short, the \emph{Ansatz} is
self-consistent.

The analysis of the right-handed Weyl equation (\ref{eq:Weyl})
proceeds in the same manner. The only subtlety is to realize \cite{PS95}
that if $\Psi_L$ has a gauge transformation employing the hermitian
representation $\hat{T}^a$, then the corresponding
$\Psi_R$ $\equiv$ $i \sigma^2 \,\Psi_L^{\;\star}$ uses the \emph{conjugate}
representation $-\hat{T}^{a {\rm T}}$ $=$ $-\hat{T}^{a \, \star}$ .
(For $SU(2)$ isodoublets, these two representations are equivalent
by unitary transformation with $i\tau^2$, which will be used later on.)
As we have seen above,
the theory (\ref{eq:AYMH}) has left-handed fermions satisfying
the Weyl equation (\ref{eq:WeylL}) in an axisymmetric gauge field
background, explicitely
\beq
   i \, \bar{\sigma}^\mu  D_\mu\, \Psi_L = 0 \; , \quad
   D_\mu \equiv \partial_\mu - i \, g\, \hat{T}^a \, W_\mu^a \; ,
\eeq
with $\hat{T}^a$ $\equiv$  $\tau^a/2$. We are therefore
interested in solving the independent equation
\beq
   i \, \sigma^\mu  \bar{D}_\mu\, \Psi_R = 0 \; , \quad
   \bar{D}_\mu \equiv \partial_\mu + i\,g\,\hat{T}^{a \, \star}\,W_\mu^a \;,
\label{eq:WeylR}
\eeq
with the same $\hat{T}^a$ $\equiv$  $\tau^a/2$ and gauge field background
$W_\mu^a$. This is accomplished by the fermion \emph{Ansatz}
(\ref{eq:PsiR}), with the resulting four equations
\begin{eqnarray}
\lefteqn{
   \left[\, \left(\partial_\tau + \frac{C_9}{2\,z} \right) \pm
    i\left(\partial_z - \frac{C_{12}}{2\,\tau} \right) \,\right]
    F_{R} \!\!\mbox{\tiny$ \begin{array}{l} \uparrow
    \\[-.5ex] \downarrow \end{array} $}-
    \left[\, \frac{1}{2\,\rho} \left( C_1 - C_4 \right) \mp
    \frac{i}{2\,\rho}\left( C_2 + C_3 \right) \,\right]
    F_{R} \!\!\mbox{\tiny$ \begin{array}{l} \downarrow
    \\[-.5ex] \uparrow \end{array} $}+
}
\nonumber\\[1ex]
\lefteqn{
    \left[\, i \left( \partial_\rho - \frac{C_5}{2\,\rho} \right) \pm
    \frac{C_6}{2\,\rho} \,\right]
    G_{R} \!\!\mbox{\tiny$ \begin{array}{c} \downarrow
    \\[-.5ex] \uparrow \end{array} $}+
    \left[\, i \left( \frac{C_{10}}{2\,\tau} + \frac{C_8}{2\,z} \right) \pm
    \left( \frac{C_{11}}{2\,\tau} - \frac{C_7}{2\,z} \right) \,\right]
    G_{R} \!\!\mbox{\tiny$ \begin{array}{c} \uparrow
    \\[-.5ex] \downarrow \end{array} $}\! = 0\;,
}
\nonumber\\[2ex]
\lefteqn{
    \left[\, \left(\partial_\tau - \frac{C_9}{2\,z} \right) \pm
    i\left(\partial_z + \frac{C_{12}}{2\,\tau} \right) \,\right]
    G_{R} \!\!\mbox{\tiny$ \begin{array}{c} \uparrow
    \\[-.5ex] \downarrow \end{array} $}
   -
    \left[\, \frac{1}{2\,\rho} \left( C_1 + C_4 \right) \pm
    \frac{i}{2\,\rho}\left( C_2 - C_3 \right) \,\right]
    G_{R} \!\!\mbox{\tiny$ \begin{array}{c} \downarrow
    \\[-.5ex] \uparrow \end{array} $}
   +
}
\nonumber\\[1ex]
\lefteqn{
    \left[\, i \left( \partial_\rho +\frac{1}{\rho} +
                      \frac{C_5}{2\,\rho} \right) \mp
    \frac{C_6}{2\,\rho} \,\right]
    F_{R} \!\!\mbox{\tiny$ \begin{array}{c} \downarrow
    \\[-.5ex] \uparrow \end{array} $}
   +
    \left[\, i \left( \frac{C_{10}}{2\,\tau} - \frac{C_8}{2\,z} \right) \mp
    \left( \frac{C_{11}}{2\,\tau} + \frac{C_7}{2\,z} \right) \,\right]
    F_{R} \!\!\mbox{\tiny$ \begin{array}{c} \uparrow
    \\[-.5ex] \downarrow \end{array} $}
   \! = 0 \;.
}
\label{eq:FGReqs}
\end{eqnarray}
Observe that for general coefficient functions $C_i$ the two sets
of equations (\ref{eq:FGLeqs}) and (\ref{eq:FGReqs}) are not equivalent
(consider, for example, the
combinations of $(C_1 \pm C_4)/2\rho$ and $(C_{10}/2\tau \pm C_8/2z)$
appearing together).

    Now turn to the Dirac equation (\ref{eq:Dirac})
in the gauge field background (\ref{eq:axialfields}), using the chiral
representation (\ref{eq:chiralrepr}) of the Dirac matrices.
This eigenvalue problem
\beq
     \left( \begin{array}{cc}
            0                 & i \bar{\sigma} \cdot D \\
            i \sigma \cdot D  & 0                     \end{array}
     \right)
\left( \begin{array}{c}\tilde{\Psi}_{R} \\ \tilde{\Psi}_{L} \end{array}\right)
= \epsilon \,
\left( \begin{array}{c}\tilde{\Psi}_{R} \\ \tilde{\Psi}_{L} \end{array}\right)
\label{eq:Diraceigenval}
\eeq
is solved by taking
\beq
\left( \begin{array}{c} \tilde{\Psi}_{R}\\ \tilde{\Psi}_{L} \end{array}\right)
=\left(\begin{array}{c} \tau^2 \,\Psi_{R}\\ \Psi_{L} \end{array}\right)\;,
\label{eq:Psitilde}
\eeq
with the wave functions on the \rhs~given by the \emph{Ans\"{a}tze}
(\ref{eq:PsiL}), (\ref{eq:PsiR}). The simple relation
\[
\tau^2 \, \tau^a \, \tau^2 = - \, \tau^{a \, \star}
\]
turns $D_\mu \tilde{\Psi}_R$  into $\tau^2 \bar{D}_\mu \Psi_R$, so that we can
use our previous results for the right-handed Weyl equation (\ref{eq:WeylR}).
The eigenvalue equation (\ref{eq:Diraceigenval}) gives then eight coupled
partial differential equations with the following structure:
\begin{eqnarray}
  \left(\partial_\tau  \mp C_9/2z \right)
  F_{\!\!\mbox{\tiny$\begin{array}{l} L\\[-.75ex] R\end{array}$}\!\!\uparrow}
  \; \cdots
&=&\mp \; \epsilon \;
  F_{\!\!\mbox{\tiny$\begin{array}{l} R\\[-.75ex] L\end{array}$}\!\!\uparrow}
\,\; , \quad
  \left(\partial_\tau  \pm C_9/2z \right)
  G_{\!\!\mbox{\tiny$\begin{array}{l} L\\[-.75ex] R\end{array}$}\!\!\uparrow}
  \; \cdots \;\;
 = \;\; \pm \; \epsilon \;
  G_{\!\!\mbox{\tiny$\begin{array}{l} R\\[-.75ex] L\end{array}$}\!\!\uparrow}
\,\; , \nonumber \\[1ex]
  \left(\partial_\tau  \mp C_9/2z \right)
  F_{\!\!\mbox{\tiny$\begin{array}{l} L\\[-.75ex] R\end{array}$}\!\!\downarrow}
  \; \cdots
&=&\pm \; \epsilon \;
  F_{\!\!\mbox{\tiny$\begin{array}{l} R\\[-.75ex] L\end{array}$}\!\!\downarrow}
\,\; , \quad
  \left(\partial_\tau  \pm C_9/2z \right)
  G_{\!\!\mbox{\tiny$\begin{array}{l} L\\[-.75ex] R\end{array}$}\!\!\downarrow}
  \; \cdots \;\;
 = \;\; \mp \; \epsilon \;
  G_{\!\!\mbox{\tiny$\begin{array}{l} R\\[-.75ex] L\end{array}$}\!\!\downarrow}
\,\; ,
\nonumber\\&&
\label{eq:redDirac}
\end{eqnarray}
where the expressions on the left-hand sides are given by the
corresponding Weyl equation results (\ref{eq:FGLeqs}), (\ref{eq:FGReqs}).

As mentioned in the Introduction, if the coefficient functions $C_i$ are
taken \cite{KW96,K93a} to belong to a \ncl~(NCL)
of bosonic field configurations
(loop parameter $\omega$ $\in$ $[-3\,\pi/2, \,3\,\pi/2]$) passing through
the vacuum ($\omega$ $=$ $\pm\, 3\,\pi/2$)
and \Istar~($\omega$ $=$ $0$),
then the eigenvalues $\epsilon (\omega)$ of (\ref{eq:redDirac})
exhibit spectral flow with an odd number of pairs of opposite levels
crossing \cite{W82,K92}.
This has been shown to follow from a 5-dimensional mod 2 \ASth~\cite{W82}.
The energy levels cross through $\epsilon = 0$ in pairs, because each
non-zero eigenvalue $\epsilon $ has a matching eigenvalue $-\epsilon $,
as follows from the anticommutation of $i\gamma^\mu D_\mu$ and $\gamma^5$.
As long as each pair of fermion zero-modes contains both chiralities,
this agrees with the standard \ASth~(see \cite{JR77} and references therein)
\beq
n_{-} - \, n_{+} = Q \;,
\label{eq:ASth}
\eeq
where $n_{\pm}$ denotes the number of normalizable zero-modes
of the isospin $\frac{1}{2}$ Dirac operator with chirality $\pm1$,
since the configurations of the NCL have topological charge $Q=0$.
Most likely, there is a single pair of levels crossing through $\epsilon = 0$
precisely at $\omega=0$ corresponding
to the \Istar~fields, which are distinguished on the NCL by
their discrete symmetry ${\rm D}_1$. If so, this implies that the reduced
Weyl equations (\ref{eq:FGLeqs}), (\ref{eq:FGReqs}) in the \Istar~background
each have a non-trivial solution.

\section{Analytic results}

    Before we look for explicit solutions of the reduced Weyl
equations (\ref{eq:FGLeqs}), (\ref{eq:FGReqs}) in the $SO(2)$
symmetric background of the instanton \Istar,
we must verify that the known results
are reproduced  in the $SO(4)$ symmetric background of the BPST instanton I.
For the self-dual gauge fields (\ref{eq:Ifields}),
(\ref{eq:Ifunctions}) the single left-handed isodoublet fermion zero-mode
\cite{tH76b} is indeed recovered
\beq
   F_{L\,\uparrow}^{\hspace{.5em}   \;\;(\,\mI\,)} =
   F_{L\,\downarrow}^{\hspace{.5em} \;\;(\,\mI\,)} = 0 \;, \quad
   G_{L\,\uparrow}^{\hspace{,5em}   \;\;(\,\mI\,)} =
-\,G_{L\,\downarrow}^{\hspace{.5em} \;\;(\,\mI\,)}
   \propto (x^2+\tilde{\rho}^{\,2})^{-3/2} \;,
\label{eq:zmI}
\eeq
whereas there is no normalizable solution for the right-handed Weyl equation.
This agrees with the index theorem (\ref{eq:ASth}) for $Q=1$ and also with the
stronger statement $n_{-}=Q$, $n_{+}=0$ valid for self-dual gauge
fields \cite{JR77}.
Similar results are obtained for the BPST anti-instanton \Ibar,
with the single right-handed isodoublet fermion zero-mode
\beq
   F_{R\,\uparrow}^{\hspace{.5em}   \;\;(\,\mIbar\,)} =
   F_{R\,\downarrow}^{\hspace{.5em} \;\;(\,\mIbar\,)} = 0 \;, \quad
   G_{R\,\uparrow}^{\hspace{,5em}   \;\;(\,\mIbar\,)} =
+\,G_{R\,\downarrow}^{\hspace{.5em} \;\;(\,\mIbar\,)}
   \propto (x^2+\tilde{\rho}^{\,2})^{-3/2} \;,
\label{eq:zmIbar}
\eeq
but without normalizable solution for the corresponding left-handed Weyl
equation. We now turn to the case of the (non-self-dual) constrained
instanton \Istar~and,
pending the complete numerical solution, study three aspects of the
problem analytically. 

\subsection{Asymptotics}

In this first subsection we investigate the asymptotic behaviour of the
left-handed fer\-mi\-onic fields in the \Istar~background.
Setting  in the \emph{Ansatz} (\ref{eq:PsiL})
\beq
   F_{L\,\uparrow} =     F_{L\,\downarrow} = 0 \;, \quad
   G_{L\,\uparrow} = -\, G_{L\,\downarrow} \;,
\label{eq:FGAnsatz}
\eeq
and using the appropriate \bcs~on the gauge field
functions $C_i$ \cite{KW96}, the Weyl equations (\ref{eq:FGLeqs})
give asymptotically three partial differential equations (PDEs)
\beq
     \left( \partial_\rho + \frac{4\,\rho}{x^2} \right) G_{L\,\uparrow} =
     \left( \partial_z + \frac{4\,z}{x^2} \right) G_{L\,\uparrow} =
     \left( \partial_\tau + \frac{4\,\tau}{x^2} \right) G_{L\,\uparrow} = 0\;,
\label{eq:3PDEs}
\eeq
together with a fourth equation vanishing trivially at infinity
\beq
    \frac{C_6}{2\,\rho} - \frac{C_7}{2\,z} - \frac{C_{11}}{2\,\tau} = 0\;.
\eeq
Taking in the \emph{Ansatz} (\ref{eq:FGL}) for $G_{L\,\uparrow}$ the functions
\beq
      k_3 = k_3(x^2) \;, \quad  k_4 = 0  \;,
\eeq
the three PDEs (\ref{eq:3PDEs}) are reduced
to a single ordinary differential equation (ODE)
\beq
     \left( \partial_{x^2} + \frac{2}{x^2} \right) k_3(x^2) = 0\;.
\eeq
This ODE gives the asymptotic solution
\beq
    \Psi_L^{\hspace{.3em} \dot{\alpha} n \;\; (\mIstar)} \propto
    \epsilon^{\dot{\alpha} n} \; x^{-4}     \;,
\label{eq:PsiLasympt}
\eeq
where the spin and isospin indices have been made explicit.
The solution (\ref{eq:PsiLasympt}) should be a reasonable
approximation for $|x|$ $>>$ $\max\,(M_W^{-1}, \mdstar /2)$,
since the theory considered has massive vector fields 
($M_W$ $=$ $\frac{1}{2}\, g\, v$), which rapidly
approach pure gauge configurations.

The index theorem (\ref{eq:ASth}), for $Q=0$, implies the existence of a
corresponding right-handed fermion zero-mode. The
asymptotic right-handed fields do not have an as simple form
as the left-handed ones and we will not give
the explicit solution here. The somewhat surprising difference
between the left- and right-handed fermion zero-modes can be traced back to
the structure of the \bcs~of the gauge fields at infinity,
as determined by the $SU(2)$ matrix
$U^\infty$ $=$ $(\hat{x}\cdot\tau)$ $i\tau^3$ $(\hat{x}\cdot\tau)^\dagger$
with the group factors in this particular order \cite{KW96}.
But the position of the angle-dependent group factors in $U^\infty$
can be interchanged by a gauge transformation. The solution $\Psi_R$
has therefore the same asymptotic behaviour ${\rm O}(x^{-4})$
as the $\Psi_L$ solution above.
These fermion zero-modes approach zero faster than the free
fermion propagator $S_{\rm F} = {\rm O}(x^{-3})$,
which implies that the effective \Istar~vertex has
derivative couplings (momentum factors), in contrast to the case of
't Hooft's effective I vertex \cite{tH76b,tH76a}.

\subsection{Symmetry axis}

In this second subsection we look for solutions of the Weyl equations
on the symmetry axis of \Istar, which in our coordinate
system coincides with the $\tau$-axis ($\rho$ $=$ $z=0$).
The general behaviour of the gauge field functions $C_i$ on the
$\tau$-axis is known \cite{KW96}, for example
\beq
C_1 = C_4 = 4\frac{\rho}{X}f_1(0,0,\tau) \;,  \quad
C_9 = 4 \frac{z}{X}f_9(0,0,\tau) \;,
\eeq
with $X^2 = \tau^2 +x_a^2$ and $x_a$ an arbitrary fixed scale parameter.
The Weyl equations (\ref{eq:FGLeqs}) for the left-handed fermions
are then reduced to
\begin{eqnarray}
\lefteqn{
    \left.
    \left(\,\partial_\tau - \frac{C_9}{2\,z} \mp i \,\partial_z \,\right) \,
    F_{L} \!\!\mbox{\tiny$ \begin{array}{l} \uparrow \\[-.5ex] \downarrow \end{array} $}
    - i \, \partial_\rho \,
    G_{L} \!\!\mbox{\tiny$ \begin{array}{c} \downarrow \\[-.5ex] \uparrow \end{array} $}
    \right|_{\rho=z=0}  = 0 \;,
}
\nonumber\\[2ex]
\lefteqn{
    \left.
    \left(\,\partial_\tau + \frac{C_9}{2\,z} \pm i \,\partial_z  \,\right) \,
    G_{L} \!\!\mbox{\tiny$ \begin{array}{c} \downarrow \\[-.5ex] \uparrow \end{array} $}
    - \frac{C_1}{\rho} \,
    G_{L} \!\!\mbox{\tiny$ \begin{array}{c} \uparrow \\[-.5ex] \downarrow \end{array} $}
    - i \,\left(\, \partial_\rho +\frac{1}{\rho} \,\right)
    F_{L} \!\!\mbox{\tiny$ \begin{array}{c} \uparrow \\[-.5ex] \downarrow \end{array} $}
    \right|_{\rho=z=0} = 0\;.
}
\label{eq:WeylLtau0}
\end{eqnarray}
Setting in the \emph{Ansatz} (\ref{eq:FGL})
\beq
     \left. k_2 = l_2 =  k_4 = l_4 = 0 \;, \quad
            k_3  = - \,l_3 \phantom{\frac{1}{1}}  \right|_{\rho=z=0}  \;,
\eeq
with $k_1(0,0,\tau)$ and $l_1(0,0,\tau)$ arbitrary but finite,
there remains a single ODE
\beq
    \partial_\tau \, k_3(0,0,\tau) + K_{+}(\tau) \, k_3(0,0,\tau) = 0 \;,
\label{eq:k3ODE}
\eeq
with
\beq
      K_{\pm}(\tau) \equiv
      \frac{2}{X} \,\left[\,2\, f_1(0,0,\tau) \pm f_9(0,0,\tau) \,\right]\;.
\label{eq:Kpm}
\eeq
The asymptotic behaviour of the gauge field functions on the $\tau$-axis
is \cite{KW96} $f_1$ $\to$ $\tau/X$ and
$f_9$ $\to$ $0$, so that (\ref{eq:k3ODE}) gives asymptotically
\beq
      k_3(0,0,\tau)  \propto \tau^{-4} \;,
\label{eq:k3asympt}
\eeq
which agrees with (\ref{eq:PsiLasympt}).
The general solution of (\ref{eq:k3ODE}) is
\beq
      k_3(0,0,\tau) =  k_3(0,0,0) \, \exp \left[\,
      - \int_{0}^{\tau} {\rm d}\tau^\prime \, K_{+}(\tau^\prime) \, \right]\;,
\label{eq:k3solution}
\eeq
in terms of the functions $f_1$ and $f_9$ on the $\tau$-axis.

    Unfortunatly, the exact functions $f_1$ and $f_9$ are not known.
We can use, instead, the approximate functions from the \ncl~construction
as summarized in the Appendix A of Ref. \cite{KW96}.
This approximation gives on the $\tau$-axis  
\beq
f_1(0,0,\tau)=f(0,\tau)\,\frac{X}{2}\,\frac{2\,\tau}{(\tau^2 - d^2/4)}\;\;,
\quad
f_9(0,0,\tau)=f(0,\tau)\,\frac{X}{2}\,\frac{d} {(\tau^2 - d^2/4)} \;\;,
\label{eq:f1f9approx}
\eeq
in terms of a single function $f=f(r,\tau)$, with $r^2$ $\equiv$ $\rho^2 + z^2$,
which has the following symmetry property and \bcs
\beq
 f(r,\tau)  =  f(r,-\tau) \;, \quad  f(0,\pm d/2) = 0 \;,\quad
 \lim_{|x| \to \infty} f(r,\tau) =  1 \;.
\label{eq:fbcs}
\eeq
A useful trial function for $f$ comes from the product of BPST
functions (\ref{eq:Ifunctions})
\beq
   f(r,\tau) = \frac{x_{+}^2}{x_{+}^2 + \tilde{\rho}^{\,2}} \;
               \frac{x_{-}^2}{x_{-}^2 + \tilde{\rho}^{\,2}} \;\; ,
\label{eq:ftrial}
\eeq
with $x_{\pm}^2 \equiv  r^2 + (\tau \pm d/2)^2 $.
The result for $K_{+}(\tau)$ as defined in (\ref{eq:Kpm}),
\beq
      K_{+}(\tau) = f(0,\tau) \, \frac{(4\,\tau + d)}{(\tau^2 - d^2/4)}\;\;,
\label{eq:K+tau}
\eeq
has no definite symmetry properties under $\tau$ $\to$ $-\tau$, nor has the
solution $k_3(\tau)$ of (\ref{eq:k3ODE}).
In fact, the solution ({\ref{eq:k3solution}) in this approximation
can be seen to have a major peak at $\tau$ $=$ $+\,d/2$ and a minor
one at $\tau$ $=$ $-\,d/2$. The relative height of this ``satellite'' peak
\beq
      \frac{k_3(0,0,-d/2)}{k_3(0,0,+d/2)} =  \exp \left[\,
      \int_{-d/2}^{+d/2} {\rm d}\tau^\prime \, K_{+}(\tau^\prime) \, \right]\;,
\label{eq:k3ratio}
\eeq
evaluated with (\ref{eq:ftrial}), (\ref{eq:K+tau}),
vanishes for $\tilde{\rho}$ $\rightarrow$ $0$
and approaches unity for $\tilde{\rho}$ $\rightarrow$ $\infty$.

The same analysis for the right-handed Weyl equation (\ref{eq:FGReqs}),
setting in the \emph{Ansatz} (\ref{eq:FGR})
\beq
     \left. m_2 = n_2 =  m_4 = n_4 = 0 \;, \quad
            m_3 = - \,n_3 \phantom{\frac{1}{1}}  \right|_{\rho=z=0} \;,
\eeq
leads to the ODE
\beq
    \partial_\tau \, m_3(0,0,\tau) + K_{-}(\tau) \, m_3(0,0,\tau) = 0 \;,
\label{eq:m3ODE}
\eeq
with $K_{-}(\tau)$ as defined by (\ref{eq:Kpm}).
The asymptotic right-handed solution $m_3(0,0,\tau)$ goes as $\tau^{-4}$,
just as for the left-handed solution (\ref{eq:k3asympt}).
But the major peak of $m_3(0,0,\tau)$, for the approximate functions
(\ref{eq:f1f9approx}), now can be seen to occur at
$\tau$ $=$ $-\,d/2$ and the minor one at $\tau$ $=$ $+\,d/2$, which
is the mirror image of the left-handed fermionic fields.

The discrete symmetry ${\rm D}_2$ of \Istar~also requires \cite{KW96}
the exact function $f_1(0,0,\tau)$ to be antisymmetric in $\tau$.
Assuming furthermore the exact function $f_9(0,0,\tau)$ to be strictly
symmetric in $\tau$, the equations (\ref{eq:k3ODE}) and (\ref{eq:m3ODE})
imply the following relation between the left- and right-handed fermion
zero-modes:
\beq
    \left| \Psi_L(0,0,\tau) \right| = \left| \Psi_R(0,0,-\tau) \right|  \;,
\label{eq:LRrelation}
\eeq
with the normalization at the origin taken equal.
Most likely, this is realized with $\Psi_L$ concentrated around
$\tau$ $=$ $+\,\mdstar/2$ and $\Psi_R$ around $\tau$ $=$ $-\,\mdstar/2$,
at least for small enough values of the scale parameter $\tilde{\rho}$.
Anyway, for non-vanishing $f_9(0,0,\tau)$ the left- and right-handed fermion
zero-modes on the $\tau$-axis are certainly different, since they solve,
under the same \bcs, different equations,
namely (\ref{eq:k3ODE}) and (\ref{eq:m3ODE}).
This difference between left- and right-handed fermion zero-modes of
\Istar~may have important physics implications, as will be discussed in
Sects. 7 and 8.

\subsection{Approximate solution}

In this third and last subsection on analytic results, we obtain an
approximation to the left-handed fermion zero-mode of \Istar~valid in the limit
$\tilde{\rho}\,v$ $\to$ $0$ and $\lambda$ $\sim$ $g^2$ $\to$ $0$. The idea is
to enlarge the classical theory (\ref{eq:AYMH}) temporarily by introducing
Wino fields $\Lambda^{\dot{\alpha} a}$ with appropriate su\-per\-symmetric
interactions.\footnote{~See Ref. \cite{ADS84} for a related discussion
of the fermion zero-modes due to the 
BPSTH instanton in genuinely su\-per\-symmetric gauge theories.
Note that also \Istar~may be relevant to dynamical supersymmetry breaking.}
The extended theory \cite{WB83} involves the vector superfield $\hat{V}$
(component fields in the
Wess-Zumino gauge: $W_\mu^a$, $\Lambda^{\dot{\alpha} a}$, $D^a$ )
and the scalar superfield $\hat{\Phi}$ (component fields:
$\Psi_L^{\dot{\alpha} n}$, $\Phi^n$, $F^n\,$),
where the auxiliary fields $D^a$ and $F^n$ can be
eliminated on-shell.

The purely bosonic instanton \Istar~remains a solution of the extended theory
in the limiting case mentioned at the beginning of this subsection.
Furthermore, a global supersymmetry transformation (parameter $\xi_\alpha$)
does not change these bosonic fields, but does generate fermionic
fields $\Psi_L$ and $\Lambda^a$, which automatically solve the field
equations, in particular \cite{WB83}
\beq
     i \, \bar{\sigma}^\mu  D_\mu\, \Psi_L =
     \sqrt{2} \, g\, \Lambda^{a\,\dagger}\, T^a \, \Phi \;.
\label{eq:SUSYeq}
\eeq
The \rhs~of (\ref{eq:SUSYeq}) vanishes by duality \cite{JR77,ADS84} close
enough to the $\tau$ $=$ $+\,\mdstar/2$ core (positive topological charge
density $q$), whereas it is only suppressed by the Higgs zero
close enough to the $\tau$ $=$ $-\,\mdstar/2$ core (negative $q$).
From the explicit supersymmetry
transformation \cite{WB83} we obtain, therefore, an approximate
solution of the left-handed Weyl equation (\ref{eq:Weyl})
in the \Istar~background
\beq
  \Psi_{L\, {\rm approx}}^{\; \,\dot{\alpha} n} =
  A_L\; i\,\sigma^{\mu\,\dot{\alpha}1}\,\xi_1\,
  \left(D_\mu \, \Phi \right)^n \;,
\label{eq:PsiLapprox}
\eeq
with $A_L=A_L(\rho,z,\tau)$ an additional amplitude factor which can become
small if necessary and with $\xi_1$ one component of a constant
spinor $\xi_\alpha$ which provides the proper normalization and mass dimension
of $\Psi_L$.

It is straightforward to verify that (\ref{eq:PsiLapprox}) respects
the continuous and discrete symmetries of the \emph{Ansatz}
(\ref{eq:PsiL}), (\ref{eq:FGL}), provided $A_L(\rho,z,\tau)$  is even in $z$.
This approximation thus reduces the eight unknown \emph{Ansatz} functions
$k_j$ and $l_j$ to one, $A_L$. Concretely, the functions $k_j$, $l_j$
are replaced by $A_L$ and specific combinations of the
known bosonic coefficient functions $C_i$, $H_j$ and their derivatives.

The approximate solution for the
right-handed fermion zero-mode has essentially the same structure as
(\ref{eq:PsiLapprox}), now with the operator $\bar{\sigma}\cdot D$ and
amplitude function $A_R$.
One expects these approximations, with suitable
amplitude functions $A_L$ and $A_R\,$, to be quite reasonable as long as
$\tilde{\rho}$ $<<$ $\mdstar /2$, so that the cores are well separated.
More importantly, the appearance of distinct composite fields
$\sigma\cdot D\,\Phi$ and $\bar{\sigma}\cdot D\, \Phi$ explains, in a way,
the intrinsic difference between the left- and right-handed fermion
zero-modes found in the previous subsection. 

This completes the discussion of the \Istar~fermion
zero-modes. We now turn to one possible application.

\section{Euclidean Green's function}

In this section and the next, the theory (\ref{eq:AYMH})
is specialized to $N_f=2$ flavors
of massless left-handed fermions. At any rate, the $N_f=1$ theory suffers from
a global (non-perturbative) gauge anomaly and is,
most likely, inconsistent \cite{W82}. Consider, then, the following euclidean
4-point Green's function evaluated by functional integration:
\begin{eqnarray}
G_{\rm E}(x_1,x_2,x_3,x_4) &\equiv&
  Z^{-1} \, \int \left[ D W \, D \Phi \,
D \Psi_1 \, D \Psi_1^{\;\dagger} \,
D \Psi_2 \, D \Psi_2^{\;\dagger} \, \right]\;
\nonumber\\[1ex]  && \times
\exp\left(\, - \,A_{\mbox {\rm \tiny YMH}}
                 \left[\,W,\,\Phi,\,\Psi_f\,\right]\,\right)
\Psi_1(x_1)\,\Psi_2(x_2)\,\Psi_1^{\;\dagger}(x_3)\,\Psi_2^{\;\dagger}(x_4)\;
\;,
\nonumber\\[1ex]  &&
\label{eq:pathintegral}
\end{eqnarray}
with the constant normalization factor $Z$,
the functional measure $[D W\,\cdots\,]$
including the necessary gauge-fixing factors and the euclidean action
functional  $A_{\mbox {\rm \tiny YMH}}$ as given by (\ref{eq:AYMH}).
Spin and isospin indices in (\ref{eq:pathintegral}) have been suppressed.
(Note that our choice of Green's function is by no means exclusive;
an alternative to (\ref{eq:pathintegral}) would be the
4-point function in terms of gauge invariant composite fields
$\Phi^\dagger \Psi_f$ and $\Psi_f^\dagger \Phi$.)

    The only hope of doing the functional integral analytically,
at least in weak coupling,  appears to be the saddle-point approximation.
But in addition to the vacuum stationary point
($W$ $=$ $\Psi_f$ $=$ $\Psi_f^{\;\dagger}$ $=$ $0$, $\Phi$ $=$ constant),
which gives rise to the standard Feynman perturbation theory \cite{F49,tH71},
there is now another stationary point to consider, namely the constrained
instanton \Istar.
The leading non-perturbative contribution to (\ref{eq:pathintegral})
follows from inserting the classical \Istar~fields and the corresponding
fermion zero-modes. Of course, there is still the integration over
the scale parameter $\tilde{\rho}$ of the solution and the
(bosonic) collective coordinates, viz.
the over-all position of the instanton \Istar, the
orientation of the \Istar~axis (in Sect. 3 taken to be along the
$\tau$-axis) and the combined global gauge and custodial \cite{SV91} $SU(2)$
transformations of the \Istar~fields.
The result is
\begin{eqnarray}
G_{\rm E}^{\; (\,\mIstar\,)}(x_1,x_2,x_3,x_4) &\propto& \,
 \int_{0}^{\infty} {\rm d}\tilde{\rho}\,
 \int_{\R^4} {\rm d}^{4}x_{\,\mIstar} \,
 \int_{S^3}  {\rm d}^{3}\Omega_{\,\mIstar} \,
 \int_{SU(2)}{\rm d}^{3} U_{\,\mIstar} \; \; n(\tilde{\rho})\;
\nonumber\\[1ex]  && \times 
 \exp\left(\,-\,A_{\mbox {\rm\tiny YMH}}^{\;\;(\,\mIstar\,)}
           (\tilde{\rho})\,\right) \;
 \Psi_1^{\;(\,\mIstar)\,}(x_1-x_{\,\mIstar})\,
 \Psi_2^{\;(\,\mIstar)\,}(x_2-x_{\,\mIstar})\,
\nonumber\\[1ex]  && \times
 \Psi_1^{\;\dagger\;(\,\mIstar\,)}(x_3-x_{\,\mIstar})\,
 \Psi_2^{\;\dagger\;(\,\mIstar\,)}(x_4-x_{\,\mIstar})\;
\;,
\label{eq:GEIstar}
\end{eqnarray}
with the factor $n$ containing the Jacobian of
the scale parameter $\tilde{\rho}$ \cite{A81,ADS84}.
The $\Psi_f$ in the integrand here correspond to the fermion zero-modes
(\ref{eq:PsiL}), (\ref{eq:FGL}), with the dependence on
$\tilde{\rho}$ suppressed,  and
the $\Psi_f^{\,\dagger}$ follow from (\ref{eq:PsiR}), (\ref{eq:FGR})
by evaluating
$\Psi_{Rf}^{\;\rm T} \,i \sigma^2$, as discussed in Sect. 5.
In principle, it is now straightforward to take the Fourier transform
of this 4-point function.

    At this point it may be of interest to compare our result with earlier
calculations, which employ an \emph{approximate} stationary point
consisting of a BPSTH instanton and anti-instanton pair \IIbar~;
see \cite{KR91} and references therein. The result corresponding to
(\ref{eq:GEIstar}) is given by Eq. (2.22) of Ref. \cite{KR91}
\begin{eqnarray}
G_{\rm E}^{\; (\,\mIIbar\,)}(x_1,x_2,x_3,x_4) &\propto& \,
  \int {\rm d}\tilde{\rho}_{\,\rm I}\;
       {\rm d}\tilde{\rho}_{\,\mIbar}\;
       {\rm d}^{4}x_{\,\rm I} \;
       {\rm d}^{4}x_{\,\mIbar}\;
       {\rm d}^{3}U_{\,\rm I} \;
       {\rm d}^{3}U_{\,\mIbar}\;\;
       n_{\,\rm I}(\tilde{\rho}_{\,\rm I}) \;
       n_{\,\mIbar}(\tilde{\rho}_{\,\mIbar}) \;
\nonumber\\[1ex]  && \times 
 \exp\left(\,-\,A_{\mbox {\rm\tiny YMH}}^{\;\;(\,\mIIbar\,)}
 \left( \,\tilde{\rho}_{\,\rm I}\,, \,\tilde{\rho}_{\,\mIbar}\,, \,
  x_{\,\rm I}-x_{\,\mIbar}\,,\,
  U_{\,\rm I}^{\phantom{1}}U_{\,\mIbar}^{-1}\,\right) \,\right)
\nonumber\\[1ex]  && \times 
 \Psi_1^{\;(\,\rm I\,)}(x_1-x_{\,\rm I})\,
 \Psi_2^{\;(\,\rm I\,)}(x_2-x_{\,\rm I})\,
 \Psi_1^{\;\dagger\;(\,\mIbar\,)}(x_3-x_{\,\mIbar})\,
\nonumber\\[1ex]  && \times 
 \Psi_2^{\;\dagger\;(\,\mIbar\,)}(x_4-x_{\,\mIbar})\; \;,
\label{eq:GEIIbar}
\end{eqnarray}
with $U_{\rm I}$ a global isospin rotation (and custodial \cite{SV91} symmetry
transformation) of the I fields as given in
(\ref{eq:Ifields}), (\ref{eq:Ifunctions}),
where $x^\mu$ is replaced by ($x^\mu - x_{\,\rm I}^\mu$) and
$\tilde{\rho}$  by $\tilde{\rho}_{\,\rm I}$, and similarly for \Ibar.
Two remarks are in order.
First, the fermion wave functions in (\ref{eq:GEIIbar})
are taken to be the standard functions (\ref{eq:zmI}) for the BPSTH instanton,
up to a global isospin rotation, and similarly (\ref{eq:zmIbar}) for the
anti-instanton.  This is certainly a reasonable approximation close to the
cores, but asymptotically (far away from both instanton and
anti-instanton) we expect the correct fermion wave functions to go as
${\rm O}(x^{-4})$, just as found in Sect 6.1.
Second, the unrestricted integration over the instanton positions
$x_{\,\rm I}$ and $x_{\,\mIbar}$ in (\ref{eq:GEIIbar})
is problematic, since an overlapping \IIbar~pair in the
attractive channel loses the fermion zero-modes altogether \cite{K92}.
The solution would be to fix the relative isospin rotation to
the repulsive channel
(for example $U_{\rm I}^{\phantom{1}} U_{\mIbar}^{-1}$ $=$ $i\tau^3$
for $x_{\rm I}^\mu-x_{\mIbar}^\mu$ $=$ $d \, \delta^{\mu 4}$),
for which case there are, most likely, still fermion zero-modes, as follows
from the discussion on spectral flow at the end of Sect. 5.
This amounts to integrating over a particular ``mountain ridge'', which has
its minimum (``mountain pass'') for a configuration in the neighbourhood of
\Istar~(see, for example, Fig. 1a of Ref. \cite{K93a}).
With our expression (\ref{eq:GEIstar})
we seem to have found (by solving the classical field equations) the
\emph{lowest possible} ``mountain pass'',
namely the exact stationary point \Istar.

Returning to the Fourier transform of the 4-point function (\ref{eq:GEIstar}),
we specialize to forward momenta $p_1=p_3$ and $p_2=p_4$.
Also, the corresponding isospin indices are matched
($n_1=n_3$ and $n_2=n_4$)                     
and averaged over.
Ultimately, this Green's function will be relevant to the \fes, which
in turn controls the total cross-section for this process.
The crucial observation now is that the expected relative shift
of the $\Psi_f$ and $\Psi_f^{\,\dagger}$ zero-modes (see Sect. 6.2) gives
rise to a non-trivial momentum phase factor in the integrand.
Assuming the average $\Psi_f$ and $\Psi_f^{\,\dagger}$ fermion wave
functions to be displaced by a distance \Dstar~along the
symmetry axis (the $\tau$-axis in our original coordinate system),
the combined exponential factor in the integrand is
\beq
      \exp\left[\, i\, E \,\mDstar \, \cos\theta -
                   A_{\mbox {\rm \tiny YMH}}^{\star} \,\right]\;,
\label{eq:phaseEucl}
\eeq
with $A_{\mbox {\rm \tiny YMH}}^{\star}$ $=$
$A_{\mbox {\rm \tiny YMH}}^{\star}(\tilde{\rho})$ 
the bosonic action of the constrained instanton \Istar,
$E$ the norm of the total momentum vector
$p_1^\mu+p_2^\mu$, and $\theta$ the angle between this vector
and the \Istar-axis.
Performing the integration over the \Istar~orientation gives the
approximate result
\begin{eqnarray}
\tilde{G}_{\rm E}^{\; (\mIstar)}(p_1,p_2,p_1,p_2) &\propto& \,
 \int_{\tilde{\rho}_{\rm \,c}}^{\infty} {\rm d}\tilde{\rho}\;
 N \;
 \exp \left(-  A_{\mbox {\rm \tiny YMH}}^{\star}  \right)\;
\frac{J_1(E\,\mDstar)}{E\,\mDstar}\;
 \left[\, 1 + {\rm O}(\tilde{\rho}\, M_W) \,\right]   \;,
\label{eq:GEfinal}
\end{eqnarray}
with $E$ the euclidean invariant $|p_1+p_2|$,
$J_1$ the Bessel function of order $1$, and
$N$ $=$ $N(\tilde{\rho},|p_1|,|p_2|)$  a  factor
containing both the Jacobian of the scale parameter $\tilde{\rho}$ and
the remaining fermion wave functions.
The precise form of the integrand in (\ref{eq:GEfinal})
depends on the details of the fermion zero-mode wave functions in
(\ref{eq:GEIstar}) and we can only give its generic structure
for small enough values of the scale parameter $\tilde{\rho}$ (as indicated
symbolically by the square bracket term in the integrand).
Generally speaking, the convergence of the integral (\ref{eq:GEfinal})
at the upper end is assured by the Higgs contribution
${\rm O}(\tilde{\rho}^2\, v^2)$ to the action,
whereas the integral at the lower end has a temporary cut-off
$\tilde{\rho}_{\rm \,c}$ (quantum corrections 
are expected \cite{tH76b} to bring in further factors of $\tilde{\rho}$,
thereby assuring the convergence).

The effective fermion shift parameter
$\mDstar$ $=$ $\mDstar(\tilde{\rho}, |p_1|, |p_2|, E)$
is implicitely defined by (\ref{eq:GEIstar}) and (\ref{eq:GEfinal}).
The value of \Dstar~is expected to be close to that of the distance
$\mdstar$ $=$ $\mdstar(\tilde{\rho})$
between the Higgs zeros of \Istar, at least for small enough values
of the scale parameter $\tilde{\rho}$ and the momenta $p_1$, $p_2$.
(For finite values of $\tilde{\rho}$, the fermion zero-mode wave functions
develop a satellite peak (\ref{eq:k3ratio}),
which tends to reduce the value of \Dstar.)
As a first approximation,
we consider $N$ and $\mDstar$ in (\ref{eq:GEfinal})
to depend only on the scale parameter $\tilde{\rho}$ and replace
\Dstar~by $\mdstar(\tilde{\rho})$.
This approximation will be used in the next section.

\section{Threshold energy}

In this last section we discuss possible
implications of the non-perturbative
euclidean 4-point Green's function found
above. The scale integration in (\ref{eq:GEfinal}), with \Dstar~replaced
by $\mdstar(\tilde{\rho})$, can be performed either
numerically or by saddle-point approximation \cite{K92}.
Making, afterwards, the analytic
continuation from euclidean to minkowskian momenta ($E$ $\to$ $i\sqrt{s}$),
results for $\sqrt{s}\,\mdstar$ $>>$ $1$ in a leading exponential factor

\beq
      \exp\left[\, \sqrt{s} \, \mdstar(\hat{\rho}) -
                A_{\mbox {\rm \tiny YMH}}^{\star}(\hat{\rho}) \,\right]\;,
\label{eq:phaseMink}
\eeq
with $\hat{\rho}$  the dominant value of $\tilde{\rho}$
in the integration (\ref{eq:GEfinal}), for which we take
$\hat{\rho}\, v$ $\sim$  $1$  (or at least $\hat{\rho}\,M_W << 1$).
The exponential suppression in (\ref{eq:phaseMink}) disappears for
large enough center-of-mass scattering energy
\beq
 (\sqrt{s})_{{\rm threshold}} \sim
\frac{ A_{\mbox {\rm \tiny YMH}}^{\star}(\hat{\rho})}{\mdstar (\hat{\rho})}=
\left(\frac{ A_{\mbox {\rm \tiny YMH}}^{\star}(\hat{\rho})}{16\,\pi^{2}/g^{2}}\right)\;
\left( \frac{ 2 \,M_W^{-1}}{\mdstar(\hat{\rho})} \right) \; \mtildeESstar   \: ,
\label{eq:Ethreshold}
\eeq
with the definition
\[ \mtildeESstar \equiv 2\, \pi\: \frac{M_{W}}{\alpha_{w}} \; ,\]
which is close to the actual energy of the \sph ~\Sstar~\cite{K93b}.
The energy $\mtildeESstar$ is approximately 20 TeV in the \ewsm.\footnote{~
\Istar -like configurations may also play a role, though perhaps a less
dramatic one, in non-chiral (vector-like) \YMths~such as QCD. }

The main uncertainty in (\ref{eq:Ethreshold}) comes
from the value of the core distance $\mdstar(\hat{\rho})$,
but $2 \,M_W^{-1}$ seems to be a reasonable first estimate.
Recall, however, that we have made a drastic simplification by setting
the effective fermion shift parameter \Dstar~equal to the distance \dstar~between
the zeros of the Higgs field.
The ultimate high-energy behaviour
remains unclear, as long as the exact fermion zero-mode
wave functions and the resulting shift parameter \Dstar~have not been
determined.  Still, the estimate (\ref{eq:Ethreshold})
may be expected to give the energy scale for non-perturbative
contributions to the Green's functions.

The particular non-perturbative contribution (\ref{eq:GEfinal})
to the forward Green's function appears to be relevant to both fermion
number ($B+L$) conserving and non-conserving \cite{tH76a} processes.
Indeed, there exists no suitable fermion number projection operator
which can be used to ``split'' the \Istar~contribution to the forward Green's
function (except, possibly, in the low-energy limit; cf. Ref. \cite{KR91}),
because of the simple fact  that in a theory with a fermion number anomaly
the fermion number is in general not well defined.
Given this new, unexpected contribution to the Green's function,
it is conceivable that chiral \YMHth~above
the ``non-perturbative threshold'' (\ref{eq:Ethreshold}) is
radically different from the standard picture based on low-order
Feynman perturbation theory.
\vspace{1\baselineskip} \par
The author thanks I. Dasgupta and P. Olesen for helpful comments
on the last two sections of this paper.


\newpage
\begin{thebibliography}{99}
\bibitem{KW96}   F. Klinkhamer and J. Weller, Nucl. Phys. B 481 (1996) 403.
\bibitem{K93a}   F. Klinkhamer, Nucl. Phys. B 407 (1993) 88.
\bibitem{W82}    E. Witten, Phys. Lett. B 117  (1982) 324.
\bibitem{K92}    F. Klinkhamer, Nucl. Phys. B 376 (1992) 255.
\bibitem{YM54}   C. Yang and R. Mills, Phys. Rev. 96 (1954) 191.
\bibitem{H66}    P. Higgs, Phys. Rev. 145 (1966) 1156.
\bibitem{A81}    I. Affleck, Nucl. Phys. B 191 (1981) 429.
\bibitem{JR77}   R. Jackiw and C. Rebbi, Phys. Rev. D 16 (1977) 1052.
\bibitem{WB83}   J. Wess and J. Bagger, \emph{Supersymmetry and Supergravity}
                 (Princeton University Press, Princeton, 1983), Chapter 7.
\bibitem{PS95}   M. Peskin and D. Schroeder, \emph{Introduction to Quantum
                 Field Theory} (Addison-Wesley, Reading, 1995), Chapter 19.
\bibitem{BPST75} A. Belavin, A. Polyakov, A. Schwartz and Yu. Tyupkin,
                 Phys. Lett. B 59 (1975) 85.
\bibitem{tH76b}  G. 't Hooft, Phys. Rev. D 14 (1976) 3432;
                 (E) D 18 (1978) 2199.
\bibitem{K93b}   F. Klinkhamer, Nucl. Phys. B 410 (1993) 343.
\bibitem{KR97}   F. Klinkhamer and C. Rupp, Nucl. Phys. B 495 (1997) 172.
\bibitem{tH76a}  G. 't Hooft, Phys. Rev. Lett. 37 (1976) 8.
\bibitem{ADS84}  I. Affleck, M. Dine and N. Seiberg,
                 Nucl. Phys. B 241 (1984) 493.
\bibitem{F49}    R. Feynman, Phys. Rev. 76 (1949) 749, 769; 80 (1950) 440.
\bibitem{tH71}   G. 't Hooft, Nucl. Phys. B 33 (1971) 173; B 35 (1971) 167.
\bibitem{SV91}   M. Shifman and A. Vainshtein, Nucl. Phys. B 362 (1991) 21.
\bibitem{KR91}   V. Khoze and A. Ringwald, Nucl. Phys. B 355 (1991) 351.
\end{thebibliography}
\end{document}